\documentclass[twocolumn,showpacs,preprintnumbers,amsmath,amssymb]{revtex4}
\usepackage{graphicx}
\usepackage{dcolumn}
\usepackage{bm}
\begin{document}
\title{Testing Bell's inequality and measuring the entanglement
using superconducting nanocircuits}
\author{Guang-Ping He$^{1}$}
\email{puaarc02@zsu.edu.cn}
\author{Shi-Liang Zhu$^{2,3}$}
\email{szhu@graduate.hku.hk}
\author{Z. D. Wang$^{2,4}$}
\email{zwang@hkucc.hku.hk}
\author{Hua-Zhong Li$^1$}
\email{puaarc@zsu.edu.cn} \affiliation{ $^{1}$Advanced Research
Center, Zhongshan University, Guangzhou 510275, China\\
$^2$Department of Physics,
University of Hong Kong, Pokfulam Road, Hong Kong, China\\
$^3$Department of
Physics, South China Normal University, Guangzhou, China\\
$^4$Department of Material Science and Engineering, University of
Science and Technology of China, Hefei, China}
\begin{abstract}
An experimental scheme is proposed to test Bell's inequality by
using superconducting nanocircuits. In this scheme, quantum
entanglement of a pair of charge qubits separated by a
sufficiently long distance may be created by cavity quantum
electrodynamic techniques; the population of qubits is
experimentally measurable by dc currents through the probe
junctions, and one measured outcome may be recorded for every
experiment. Therefore, both locality and detection efficiency
loopholes should be closed in the same experiment. We also propose
a useful method to measure the amount of entanglement based on the
concurrence between Josephson qubits. The measurable variables for
Bell's inequality as well as the entanglement are expressed in
terms of a useful phase-space Q function.
\end{abstract}
\pacs{03.65.Ud, 85.25.Cp, 85.35.-p} \maketitle

\section{Introduction}

Recently, with the development of experimental techniques and the
growing interest in quantum information, more and more attention
has been devoted to experimentally testing the violation of Bell's
inequality\cite{Bell}, as well as measuring the amount of
entanglement of entangled particles. Entanglement of particles, an
idea introduced in physics by the famous Einstein-Podolsky-Rosen
(EPR) gedanken experiment \cite{EPR}, is one of the most
strikingly nonclassical features of quantum theory. In quantum
mechanics, particles are called entangled if their states can not
be factored into single-particle states. This inseparability leads
to a stronger correlation between entangled quantum systems than
classical ones. Since Bell's pioneering work\cite{Bell} that EPR's
implication to explain the correlations using hidden parameters
would contradict the predictions of quantum physics, a number of
experimental tests have been
performed\cite{Aspect,Shih,Ou,Tittel,Weihs,Rowe}. Many of these
experiments \cite{Aspect,Shih,Ou,Tittel,Weihs} have been done by
using photons to prepare EPR pairs. Very recently, trapped atoms
were also used in the experiment\cite{Rowe} for testing Bell's
inequality to raise the efficiency when reading out the state. The
violation of Bell's inequality may be considered as a
manifestation of the irreconcilability of quantum mechanics and
``local realism'', and all recent experiments have agreed with the
predictions of quantum mechanics.

Nevertheless, two important experimental loopholes mean that the
evidence reported in the previous experiments was
inclusive\cite{loophole,Grangier}. The first of these loopholes is
the so-called locality loophole: whenever measurements are
performed on two spatially separated particles, any possibility of
signals propagating with a speed equal to or less than the
velocity of light between the two parts of the apparatus must be
excluded. This loophole was closed in the experiments reported in
Refs.\cite{Tittel,Weihs}. The second one is referred to as the
detection-efficiency loophole, which argues that in most optical
experiments, only a very small fraction of the particles generated
are actually detected. So it is possible that for each
measurement, the statistical sample provided by the detector is
biased. Since improving the detection efficiency in experiments
with pairs of entangled photons is found to be more difficult than
expected, closing this loophole experimentally is achieved by
using two massive entangled trapped ions\cite{Rowe}, where the
states are easier to be detected than those of photons. But the
experiment\cite{Rowe} does not close the locality loophole. To
close both loopholes in the same experiment remains to be a big
challenge at present\cite{Grangier}.

In this paper, we propose a scheme to test Bell's inequality in
the Clauser, Horne, Shimony and Holt (CHSH)\cite{Clauser} type and
to measure the entanglement in superconducting nanocircuits. At
first glance, the possibility of testing Bell's inequality in this
system may seem to be a trivial generalization of the
corresponding tests by using
photons\cite{Aspect,Shih,Ou,Tittel,Weihs} or trapped
ions\cite{Rowe}. However, we show that the experiment proposed
here has its own advantages. First, it is possible that both
loopholes mentioned above may be closed in the same experiment.
Thus, the loophole-free experiment proposed here may lead to a
full logically consistent rejection of any local realistic
hypothesis\cite{Grangier}. Recently, very promising development
was reported for Josephson-junction qubits under
control\cite{Josephson,Nakamura,Pashkin,Zhu_prl2002}. The charge
state in a superconducting box may be considered as a qubit
system. We first address how to prepare a pair of entangled charge
qubits in a sufficiently long distance\cite{Zhu}, which fully
enforces the requirement for strict relativistic separation
between measurements. Contrary to the experiments using photons,
where many photon pairs are missed, the charge states in the
superconducting boxes may be detected in every measurement, thus
the data in Bell's experiments are obtained using the outcome of
every experiment, thereby no fair-sampling hypothesis is required.
Consequently, {\sl both the locality loophole and the efficiency
loophole should be closed in the same experiment proposed here}.
Second, quantum mechanics violates Bell's inequality for a state
of $N$-qubit system by an amount that grows exponentially with
$N$\cite{Mermin}. The nature of being solid-state based makes
Josephson junction system large-scalable, that is, the
Greenberger-Horne-Zeilinger (GHZ) state [the maximally entangled
states of $N\ (>2)$ particles ]\cite{GHZ} may be achieved, in
principle, in superconducting nanocircuits. However, due to
experimental techniques, it is very hard to obtain many-particle
entangled states in realistic experiments with photons or trapped
atoms. Last, but not the least, the Josephson junction system
proposed here is a mesoscopic system. Comparing with the systems
consisting of a small number of microscopic particles, such as
photons or trapped ions, where quantum entanglement is generally
believed to exist, the superconducting box considered here
involves a huge number of Cooper pairs. To test the nonlocality of
a system containing a large number of particles is attractive for
research on the border between classical and quantum physics.
Furthermore, since the CHSH type of Bell's inequality is not very
efficient for demonstrating nonlocality and all entangled states
would violate a kind of Bell's inequality\cite{Capasso}, detection
of the amount of entanglement is also proposed here.

This paper is organized as follows. In Sec. II, we demonstrate
that the three basic ingredients required by Bell's experiment are
experimentally feasible in Josephson charge qubit systems. In Sec.
III, an experimental scheme for testing the Bell's inequality is
proposed. In Sec. IV, measuring the entanglement based on
concurrence in the Josephson junction systems is studied. The
paper ends with a brief summary.

\section{Entangled states in Josephson Junctions}

A Bell's experiment suggested by CHSH \cite{Clauser} consists of
three basic ingredients\cite{Rowe}. The first one is the
preparation of a pair of entangled particles in a repeatable way,
with the two particles being separated with a sufficiently long
distance. Second, each particle can be manipulated by any rotation
operations (single-qubit gates). Finally, a classical property
with two possible outcomes may be detected for each particle. We
now show that all these three ingredients are feasible in
superconducting nanocircuits.

\begin{figure}[htbp]
\centering
\includegraphics[width=8cm]{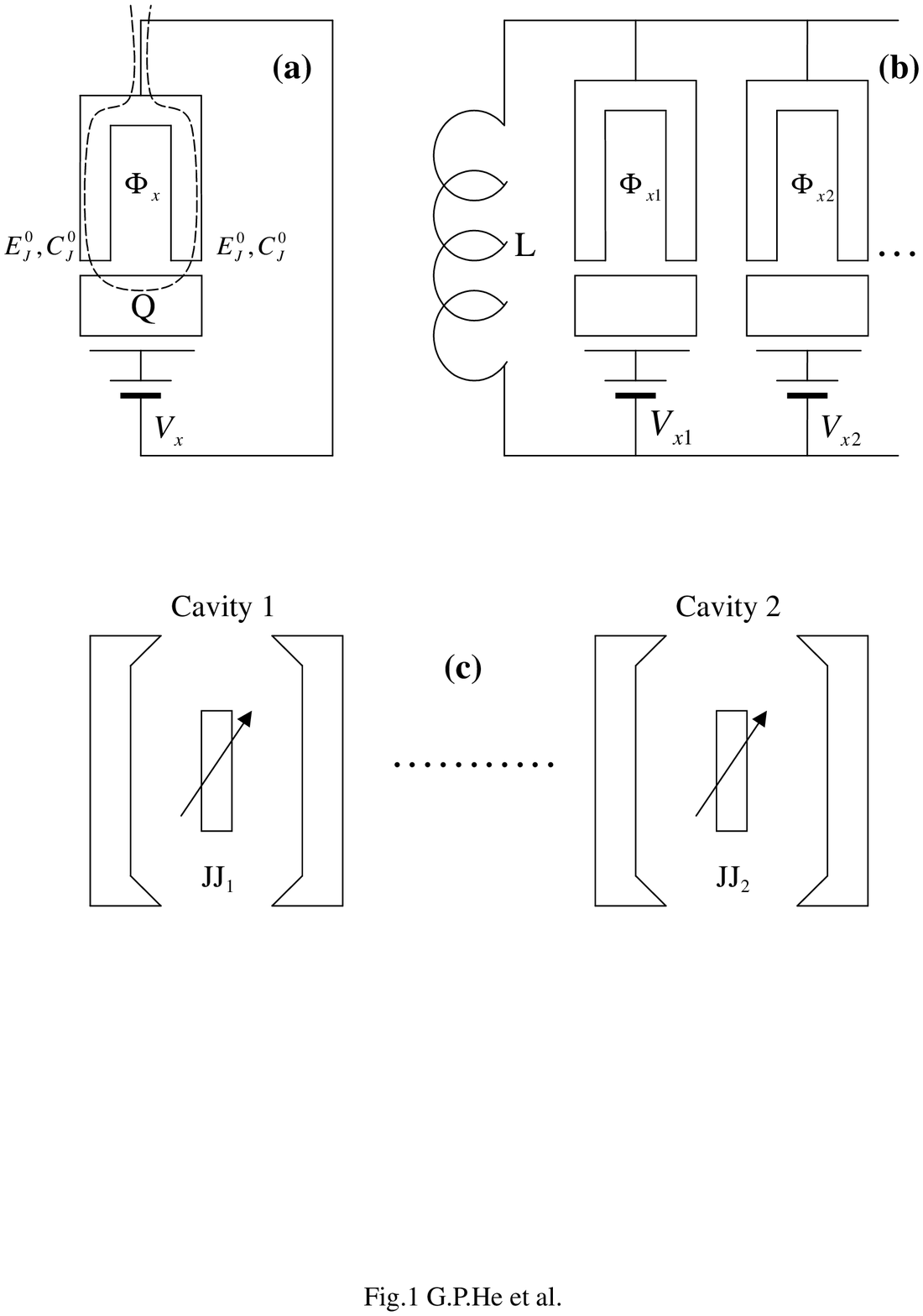}
\vspace{-3cm} \caption{Josephson qubits system. (a) A single
Josephson qubit (Ref.[12]). A superconducting island is coupled by
two Josephson junctions (each with capacitance $C_{J}^{0}$ and
Josephson coupling energy $E_{J}^{0}$) to a superconducting lead,
and through a gate capacitor to a voltage source $V_{x}$. This
dc-SQUID is tuned by external flux $\Phi_{x}$, which is controlled
by the current through the inductor loop (dashed line). (b) A
series of Josephson qubits coupled by the LC-oscillator mode
(Ref.[12]). (c) Schematic picture of quantum transmission between
two Josephson junction charge qubits in cavities connected by a
quantum channel.} \label{fig1}
\end{figure}

{\sl The charge qubits in Josephson junctions}. The systems we
considered are shown in Fig. 1. A single Josephson-junction qubit
consists of a superconducting electron box formed by a symmetric
superconducting quantum interference device (SQUID) with Josephson
couplings $E_J^0$, pierced by a magnetic flux $\phi_x$ and subject
to an applied gate voltage $V_x=2en_x/C_x$ [$2en_{x}$ is the
offset charge, see Fig.1(a)]. In the charging regime [where
$E_J^0\ll E_{ch}$ with $E_{ch}=e^2/2(C_x+2E_J^0)$ being the
single-electron charging energy] and at low temperatures, the
system behaves as an artificial spin-1/2 particle in a magnetic
field, and the Hamiltonian may be expressed as \cite{Josephson}
\begin{equation}
\label{single}
H=-\frac{1}{2}E_{J}\sigma _{x}-E_{ch}(1-2n_{x})\sigma _{z},
\end{equation}
where $E_{J}=2E_{J}^{0}\cos (\pi \phi _{x}/\phi _{0})$ with
$\phi_0=\pi\hbar/e$ is a tunable effective Josephson coupling
parameter, whose value can be controlled by external flux $\phi
_{x}$. $\sigma _{x,z}$ are the Pauli matrices. In
Eq.(\ref{single}), we have chosen that charge states $n=0$ and
$n=1$ (here $n$ is the number of excess Cooper-pair charges on the
box) correspond to spin basis states $|\uparrow\rangle\equiv
\left(
\begin{array}{c}
1\\
0
\end{array}\right)$
and
$|\downarrow\rangle\equiv \left(
\begin{array}{c}
0\\
1
\end{array}\right)$,
respectively.
A series of $N$ such qubits may be coupled through an inductor $L$
[see Fig.1(b)]. An
effective interaction is given by\cite{Josephson}
\begin{equation}
H_{int}=-\sum\limits_{i<j}\frac{E_{J}^{(i)}E_{J}^{(j)}}{E_{L}}\sigma
_{y}^{(i)}\cdot \sigma _{y}^{(j)},
\end{equation}
where $E_{L}=[\phi_0^2/(\pi^2 L)](1+2C_J^0/C_x)^2$ and $\sigma
_{y}^{(j)}$ is the $y$-component of Pauli matrices of the $j$th
qubit. Then the total Hamiltonian of the system is equivalent to
\begin{equation}
\label{H_system}
H=\frac{1}{2}\sum\limits_{i=1}^{N}
(\sigma _{x}^{(i)} B^{(i)}_x+\sigma _{z}^{(i)} B^{(i)}_z)
+\sum\limits_{i<j} J^{(ij)}\sigma
_{y}^{(i)} \sigma _{y}^{(j)},
\end{equation}
where $B_x^{(i)}\equiv -E_{J}(\phi_x^{(i)})$, $B_z^{(i)}\equiv
-2E_{ch}(1-2n_x^{(i)})$, and $J^{(ij)} \equiv -E_{J}^{(i)}
E_{J}^{(j)}/E_{L}$. It is worth pointing out that all parameters
in Eq. (\ref{H_system}) are experimentally controllable by
external classical variables $\phi_x^{(i)}$ and $n_x^{(i)}$.
Thereby, any entangled state as well as single qubit gates
required in Bell's experiment is feasible.

{\sl The preparation of a pair of entangled qubits}. For
simplicity, but without loss of generality, we consider the
two-qubit case. By taking $B_z^{(i)}=0$, $B=-E_{J}^{(i)}/2$ for
each single Josephson qubit, and $J\equiv -E_{J}^2 /E_{L}$, the
Hamiltonian may be rewritten as
\begin{equation}
\label{S1}
H=B(\sigma _{x}^{(1)}+\sigma _{x}^{(2)})+J\sigma_{y}^{(1)} \sigma _{y}^{(2)}.
\end{equation}
The exact solution may be obtained when parameters $B$\ and $J$
are time-independent. Note that any entangled state may be
generated even if the initial state is a product state. For
example, in terms of computational basis $\{ |00\rangle,
|01\rangle, |10\rangle, |11\rangle \}$, with the initial state
being chosen as $| \psi (t=0) \rangle = |00\rangle $, the state of
the system at time $t$\ is found to be
\begin{equation}
| \psi (t) \rangle =\left[
\begin{array}{c}
\label{state}
\frac{1}{2}e^{-iJt}+\frac{\alpha -J}{4\alpha }e^{-i\alpha t}+\frac{\alpha +J%
}{4\alpha }e^{i\alpha t} \\
-i\frac{B}{\alpha }\sin (\alpha t) \\
-i\frac{B}{\alpha }\sin (\alpha t) \\
-\frac{1}{2}e^{-iJt}+\frac{\alpha -J}{4\alpha }e^{-i\alpha t}+\frac{\alpha +J%
}{4\alpha }e^{i\alpha t}
\end{array}
\right],
\end{equation}
where $\alpha \equiv \sqrt{4B^{2}+J^{2}}$.
Besides, when $t=n\pi /\alpha $, $%
J=(m+1/2)\alpha /n$ ($m$, $n$ are both integers and $n\neq 0$),
we have
\begin{equation}
|\psi (t)\rangle =\frac{(-1)^{n}-i(-1)^{m}}{2}|00\rangle +\frac{%
(-1)^{n}+i(-1)^{m}}{2}|11\rangle .  \label{bell}
\end{equation}
This is the maximally entangled state for a pair of qubits [as the
concurrence of this state defined in Eq. (\ref{Concurrence}) below
is $1$], and thus, four Bell's states may be derived from it by
simply rotating one of the qubits. In the following discussions we
assume $|\psi \rangle $ in Eq. (\ref{state}) as our starting point
for the test.

{\sl The entanglement between two distant Josephson junction
qubits} may be created by the cavity quantum electrodynamic (QED)
techniques. A simple configuration of quantum transmission between
two nodes consists of two Josephson charge qubits $1$ and $2$
which are strongly coupled to their respective cavity modes with
the same frequency $\nu$, as shown in Fig. 1(c). The Hamiltonian
describing the interaction of the qubit with the cavity mode
is\cite{Zhu}
\begin{eqnarray}
\nonumber H_j&& =  \hbar \nu (a^\dagger_j a_j+\frac{1}{2})
+E_{ch}(2n^{(j)}_x-1)\sigma_j^z \\
\label{Cavity} && -\frac{1}{2} E_J(\phi_j) (e^{-i[g
(a+a^\dagger)]} \sigma_j^{+}+H.c) \ \ (j=1,2),
\end{eqnarray}
where $ a^\dagger_j$ and $a_j$ are the creation and annihilation
operators, respectively, for cavity mode $j$,
$\sigma_j^{\pm}=(\sigma_j^x \pm i\sigma_j^y)/2$, and $g$ is the
coupling constant between the junctions and the cavity. Based on
this kind of coupling and following the method described in Ref.
\cite{Cirac}, we may transfer the quantum state in one qubit to
another separated far away. Moreover, by using quantum
repeaters\cite{Briegel}, any long-distance entanglement may be
realized, at least in principle. Therefore, a possible scenario to
generate an entangled state of long-distant qubits required by
nonlocally testing Bell's inequality is as follows: we may first
create an initial entangled state [as Eq. (\ref{state})] of two
charge qubits in one node, and then transfer the state of one of
the pair to another node by the cavity QED technique.

{\sl A universal set of single-qubit gates} is feasible in the
systems. The coupling between charge qubits should be switched off
by setting $J=0$ in single-qubit gates. Parameters $\phi_x$ and
$n_x$ in Eq.(\ref{single}) are experimentally controllable. By
assuming $\phi_x=\phi_0/2$ and $n_x$ time-independent, the
evolution operator is derived as
\begin{equation}
\label{Uz}
U_z (\theta_z) = \exp(-i \theta_z \sigma_z/2)
\end{equation}
with $\theta_z =2 E_{ch}(1-2n_x) t/\hbar$. Similarly,
by assuming  $n_x=1/2$ and $\phi_x$ time-independent,
we have
\begin{equation}
\label{Ux} U_x (\theta_x) = \exp(-i \theta_x \sigma_x/2)
\end{equation}
with $\theta_x =2 E_J(\phi_x) t/\hbar$.
The gates described by Eqs.(\ref{Uz})
and (\ref{Ux}) are a well-known
universal set of single-qubit gates:
any unitary rotation can be decomposed into
a product of successive gates
in this set.
The gates described by Eqs.(\ref{Uz}) and (\ref{Ux})
may be referred to as dynamic gates\cite{Zhu_pra2003}.
It is worth pointing out that a universal set
of quantum gates in this system may also
be realized by using pure geometric phases\cite{Zhu_prl2002}.

{\sl The detection method}. The population of qubits in states
$|0\rangle$ or $|1\rangle$ may be experimentally measured  by the
dc currents through the probe junctions\cite{Nakamura}. Putting a
probe junction in each qubit, as described in Ref.\cite{Nakamura},
the measurable dc currents through the probe junction are
generated by the following process: $| 1 \rangle $ emits two
electrons to the probe, while $|0\rangle$ does nothing.
Consequently, a classical property with two possible outcomes as
required by Bell's experiments, may be detected. The advantage of
the above detection technique lies in that a measured outcome may
be recorded for every experiment, thereby closing the detection
loophole in the same experiment.

All in all, the three basic ingredients required by CHSH type
Bell's experiment are, in principle, feasible in superconducting
nanocircuits presented here.

\section{Testing Bell's inequality}

The detection method addressed above provides a convenient way to
experimentally measure the population of the Josephson qubit in
state $| 0\rangle $ or $| 1\rangle $, and hence is sufficient for
testing the Bell inequality\cite{Brif}. We show here that the CHSH
combination\cite{Clauser} can be presented by a useful phase-space
distribution function Q for the qubits, which can be calculated
through the probabilities of finding certain qubits in state $|
0\rangle $ or $| 1\rangle $. Assuming that two distant qubits
described by Eq. (\ref{state}) have been created, and each qubit
can be manipulated by unitary operators $U_{x,z}$ in
Eqs.(\ref{Uz}) and (\ref{Ux}), a new state given by
\begin{equation}
\label{final}
| \psi ({\mathbf{n}}_{1},{\mathbf{n}}_{2})\rangle
=g_{1}^{+}({\mathbf{n}}_{1})g_{2}^{+}({\mathbf{n}}_{2})| \psi \rangle
\end{equation}
is derived by rotating separately each qubit in $|\psi\rangle$
describe by Eq. (\ref{state}). Evolution operator
$g_j({\mathbf{n}}_{j})=U_z(\phi_j) U_x(\theta_j)$ with a unit
vector ${\mathbf{n}}_{j}=(\sin \theta_{j}\cos \phi_{j}, \sin
\theta_{j}\sin \phi_{j},\cos \theta _{j})$ and $j$ $(=1,2)$
denoting qubit $j$. When measuring state $| \psi
({\mathbf{n}}_{1},{\mathbf{n}}_{2})\rangle $, the probability to
find the $j$th qubit in state $|0\rangle$ is
\begin{equation}
\label{Qi}
Q_{j}({\mathbf{n}}_{j})=\langle 0_j | Tr_{i\neq j}
\{ | \psi ({\mathbf{n}}_{1},{\mathbf{n}}_{2})\rangle \langle
\psi ({\mathbf{n}}_{1},{\mathbf{n}}_{2}) | \} | 0_j\rangle,
\end{equation}
and the probability to find both qubits in state $|0\rangle$ is
\begin{equation}
\label{Q12}
Q_{12}({\mathbf{n}}_{1},{\mathbf{n}}_{2})=| \langle 0_1 |\langle 0_2|
\psi ({\mathbf{n}}_{1},{\mathbf{n}}_{2}) \rangle |^{2}.
\end{equation}
On the other hand, we may write \begin{eqnarray*} |
{\mathbf{n}}_{j}\rangle &=&g_j ({\mathbf{n}}_{j}) | 0_j\rangle \\
&=&\cos(\theta_{j}/2)e^{-i\phi _{j}/2} | 0_j\rangle -i\sin
(\theta_{j}/2)e^{i\phi _{j}/2} | 1_j\rangle,
\end{eqnarray*}
which can be understood as the qubit state in
the phase space.
Then we find that Eq.(\ref{Qi}) can be rewritten as
\begin{equation}
Q_{j}({\mathbf{n}}_{j})=\langle {\mathbf{n}}_{j}| \rho _{j}
| {\mathbf{n}}_{j}\rangle,
\end{equation}
which is just the Q function for the $j$th qubit, where $\rho
_{j}=Tr_{i\neq j}\{ |\psi \rangle \langle \psi | \}$ denotes the
reduced density matrix of the $j$th qubit. Equation (\ref{Q12})
can also be rewritten as
\begin{equation}
Q_{12}({\mathbf{n}}_{1},{\mathbf{n}}_{2})=| \langle {\mathbf{n}}_{1}
|\langle {\mathbf{n}}_{2} | \psi \rangle |^{2},
\end{equation}
which is just the joint Q function for the system of two qubits. Thus
the CHSH combination is given by
\cite{Brif,Banaszek}
\begin{eqnarray}
\nonumber
\Gamma =& & Q_{12}(0,0)+Q_{12}({\mathbf{n}},0)+Q_{12}
(0,{\mathbf{n}}^{\prime })
-Q_{12}({\mathbf{n}},{\mathbf{n}}^{\prime })
\\
& & -Q_{1}(0)-Q_{2}(0),
\label{CHSH}
\end{eqnarray}
which  must satisfy inequality $-1\leq \Gamma \leq 0$ for local
theories.

\begin{figure}[htbp]
\centering
\includegraphics[width=8cm]{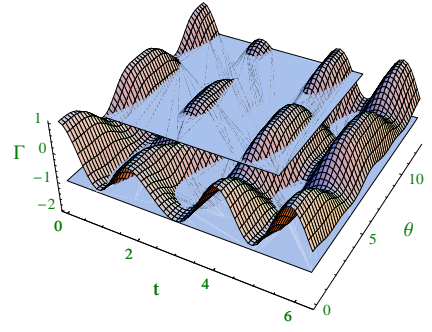}
\vspace{-5cm} \caption{Dimensionless $\Gamma$ as a function of $t$
$(1/\alpha)$ and $\theta $ when $B=J=1$ and $\phi=\protect\phi
^{\prime }=0$. The two plates cut $\Gamma $ at $0$ and $-1$,
respectively. } \label{fig2}
\end{figure}

We now consider a two-qubit system. Substituting the state
described by Eq.(\ref{state}) into Eq. (\ref{CHSH}), we have
\begin{eqnarray}
\Gamma  &=&-\sin ^{4}\frac{\theta }{2}+\cos \alpha t\cos Jt-\frac{J}{\alpha }%
\sin \alpha t\sin Jt  \nonumber \\
&&+\frac{1}{4}(\cos \alpha t\sin Jt+\frac{J}{\alpha }\sin \alpha t\cos
Jt)\sin ^{2}\theta \sin (\phi _{1}+\phi _{2})
\nonumber \\
&&-\frac{B^{2}}{\alpha ^{2}}\sin ^{2}\alpha t(2-4\sin ^{4}\frac{\theta }{2}%
+\sin ^{2}\theta \cos \phi _{1}\cos \phi _{2})
\nonumber \\
&&+\frac{B}{\alpha }\sin \alpha t\sin \theta \sin ^{2}\frac{\theta }{2}[\sin
Jt(\sin \phi _{1}+\sin \phi _{2})
\nonumber \\
&&+\cos \alpha t(\cos \phi _{1}+\cos \phi
_{2})].
\end{eqnarray}
For simplicity, but without loss of generality,
we here have chosen $\theta_{1}=\theta _{2}=\theta $
for $| {\mathbf{n}}_{1}\rangle $ and
$|{\mathbf{n}}_{2}\rangle $ in the calculation. In Fig.2, we plot
$\Gamma $\ as a function of $t$\ and $\theta $ when $%
B=J=1$\ and $\phi _{1}=\phi _{2}=0$. It is seen clearly that the
violation of Bell's inequality for both $\Gamma <-1$\ and $\Gamma
>0$ may appear in this system. In fact, with other choices for
parameters $B$, $J$, $\phi _{1}$, and $\phi _{2}$, function
$\Gamma $\ will still have a similar shape and the violation can
easily be found. Therefore, the system presented here may be a
promising candidate for testing Bell's inequality.

\section{Measuring the entanglement of formation}

On the other hand, it is well-known that Bell's inequality is
not very efficient for
demonstrating nonlocality, and
a better parameter to characterize nonlocality should be
the amount of entanglement.
Thus it is also highly desirable to develop a feasible method to measure
the latter. We now show that
the entanglement based on concurrence
can also be represented by the phase-space Q
function and thus can be experimentally detected.

If we write a state\ in the form as
$$
| \psi \rangle =a_{0} |00\rangle +a_{1}|
01\rangle +a_{2}| 10\rangle +a_{3} | 11\rangle,
$$
then concurrence $C$ of state $|\psi\rangle$ is defined
as\cite{Wootters}
\begin{equation}
\label{Concurrence}
C^{2}(|\psi\rangle )=|\langle \psi | \sigma _{y}\otimes \sigma
_{y}| \psi ^{\ast }\rangle | ^{2}=4|
a_{0}a_{3}-a_{1}a_{2}| ^{2},
\end{equation}
and the amount of entanglement can be expressed as
\begin{eqnarray}
\nonumber
E(|\psi\rangle ) &=&-\left( \frac{1+\sqrt{1-C^{2}}}{2}\right)
\log _{2}\left( \frac{1+\sqrt{1-C^{2}}}{2}\right)
\\
\label{E_concurrence}
&&-\left( \frac{1-\sqrt{1-C^{2}}}{2}\right) \log _{2}
\left( \frac{1-\sqrt{%
1-C^{2}}}{2}\right) .
\end{eqnarray}
Substituting Eq. (\ref{state}) into Eq. (\ref{Concurrence}),
the concurrence of the
state in Eq. (\ref{state}) is derived as
\begin{eqnarray}
\nonumber
C^{2}(|\psi\rangle )= & &\frac{J^{4}}{\alpha ^{4}}\sin ^{4}(\alpha t)+\frac{J^{2}}{%
4\alpha ^{2}}[\sin ^{2}(2\alpha t)-8\sin ^{2}(\alpha t)\sin ^{2}(Jt)]
\\
\label{theor}
& & +\frac{J}{2\alpha }\sin (2\alpha t)\sin (2Jt)+\sin ^{2}(Jt).
\end{eqnarray}

Let $P_{0}$, $P_{1}$, $P_{2}$ and $P_{3}$ denote the four
probabilities associated with outcomes ($\left| 00\right\rangle $,
$\left| 01\right\rangle $, $\left| 10\right\rangle$, and $\left|
11\right\rangle $) when measuring $\sigma _{z}\otimes I$\ and
$I\otimes \sigma _{z}$, and $P_{++}$, $P_{+-}$, $P_{-+}$, and
$P_{--}$ denote the four
corresponding probabilities when measuring $\sigma _{z}\otimes I$\ and $%
I\otimes \sigma _{x}$. The concurrence in
Eq.(\ref{Concurrence}) is found to satisfy\cite{entangle}

\begin{equation}
\label{concurrence}
C^{2} =4 [P_{1}P_{2}+P_{0}P_{3}-2\sqrt{P_{0}P_{1}P_{2}P_{3}} \cos(\alpha+\beta)],
\end{equation}
with
\begin{eqnarray*}
\cos\alpha &=& \frac{2P_{++}-P_{0}-P_{1}}{2\sqrt{P_{0}P_{1}}},  \\
\cos\beta  &=&  \frac{2P_{-+}+P_{0}+P_{1}-1}{2\sqrt{P_2 P_3 }}.
\end{eqnarray*}

The concurrence of a pure two-qubit state
may be measured by detection of the Q function
defined in Eqs.(\ref{Qi}) and (\ref{Q12})
since all variables in Eq. (\ref{concurrence}) are determined
by them. By choosing
\begin{eqnarray*}
{\mathbf{n}}_{a} &=&(0,0,-1),\\
{\mathbf{n}}_{b} &=&(0,1,0), \\
{\mathbf{n}}_{c} &=&(0,-1,0),
\end{eqnarray*}
which correspond to
\begin{eqnarray*}
g^{+}({\mathbf{n}}_{a}) &=& e^{i\pi \sigma _{x}/2}, \\
g^{+}({\mathbf{n}}_{b}) &=& e^{i\pi \sigma _{x}/4}e^{i\pi \sigma _{z}/4},\\
g^{+}({\mathbf{n}}_{c}) &=& e^{i\pi \sigma_{x}/4}e^{-i\pi \sigma _{z}/4},
\end{eqnarray*}
we find that the probabilities appeared in Eq.(\ref{concurrence})
are related to the Q function by
\begin{eqnarray*}
&& P_{0} = Q_{12}(0,0),
\ \ P_{1} = Q_{12}(0,{\mathbf{n}}_{a}),\\
&& P_{2} = Q_{12}({\mathbf{n}}_{a},0),
\ \  P_{3} = Q_{12}({\mathbf{n}}_{a},{\mathbf{n}}_{a}),\\
&& P_{++} = Q_{12}(0,{\mathbf{n}}_{b}),
\ \ P_{+-} = Q_{12}(0,{\mathbf{n}}_{c}),\\
&& P_{-+} = Q_{12}({\mathbf{n}}_{a},{\mathbf{n}}_{b}),
\ \ P_{--} = Q_{12}({\mathbf{n}}_{a},{\mathbf{n}}_{c}).
\end{eqnarray*}
Thus the concurrence as well as the amount of entanglement can be
deduced by detecting the probabilities of qubits in state
$|0\rangle$.

\begin{figure}[htbp]
\centering
\includegraphics[width=8cm]{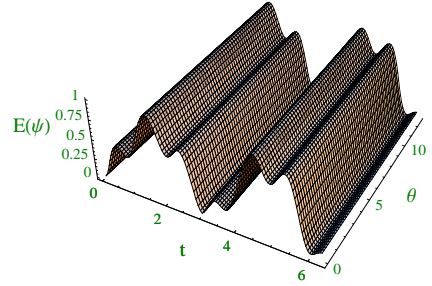}
\vspace{-5cm} \caption{Dimensionless $E(|\psi\rangle)$ as a
function of $t$ $(1/\alpha)$ and $\theta $. The parameters are the
same as those in Fig.2.} \label{fig3}
\end{figure}

A theoretical result for the amount of entanglement calculated
from Eq.(\ref{theor}) is plotted in Fig.3. We can see that
$E(|\psi\rangle)$ does not dependent on $\theta $. This
illustrates that unlike the violation of the Bell's inequality,
the entanglement will not be affected by local transformations.
Therefore, the amount of entanglement for the states given by
Eqs.(\ref{state}) and (\ref{final}) are the same. This also
implies that in the sense of characterization of the nonlocal
properties, the amount of entanglement defined by Eq.
(\ref{E_concurrence}) is better than the CHSH combination
described by Eq. (\ref{CHSH}). Thus it is desirable to work out a
feasible method to measure the amount of entanglement.

\section{Conclusions}

In conclusion, we proposed a scheme to test
Bell's inequality and to measure the entanglement of two charge qubits
in superconducting nanocircuits.
We demonstrated that the parameters for
these experiments are
determined from the Q functions in phase space,
which can be measured by the dc currents through the
probe junctions.
The outcome for every experiment may be recorded, and thus
the issue of detection efficiency is replaced
by detecting accuracy\cite{Rowe}.
By using the cavity QED technique,
the entangled state of two charge qubits separated far away
may be created. Consequently,
it is quite possible that both of the locality loophole
and the efficiency loophole
can be closed in the same experiment proposed here.

Finally, we wish to make a few remarks on the   difficulties of
experimental implementation of the scheme. (1) Designing the
cavity QED to couple the charge qubits is necessary. The
entanglement between two charge qubits was demonstrated in a
recent experiment\cite{Pashkin}, and a few quantum phenomena, such
as stimulated emission and amplification in Josephson junction
arrays within the same high-Q oscillators, were also reported
\cite{Barbara}. However, it is still awaited to make the entangled
state between the charge qubit and the single cavity mode. (2) The
distance between two cavities is enforced by the strict
relativistic separation. Since
 the measurement time
is $64$ns  in the experiment reported in Ref.\cite{Pashkin} or
may, in principle, be even shorter with one order of the
magnitude\cite{Josephson}, it is estimated that the two cavities
would have to be physically separated by $2 \sim 20 m$, which can
be realized  with current technology if the cavities are connected
by a (microwave) transmission fibre\cite{Zhu,Cirac}; the
transmission of entangled state with the distance of a few
kilometers was already realized in quantum
telepotation\cite{Marcikic}. However, it is still very subtle and
challenging to couple the (microwave) fibre to the cavity QED
experimentally.

\section{Acknowledgments}

This work was supported in part by the RGC grant of Hong Kong
under Grant No. HKU7114/02P and the URC fund of HKU. G. P. H. was
supported in part by the NSF of Guangdong Province under Contract
No. 011151 and the Foundation of Zhongshan University Advanced
Research Centre under Contract No. 02P2. S. L. Z was supported in
part by SRF for ROCS, SEM, the NSF of Guangdong province under
Grant No. 021088, and the NNSF of China under Grant No. 10204008.

\end{document}